\documentstyle[12pt]{article}
\advance \topmargin by -\headheight
\advance \topmargin by -\headsep
     
\evensidemargin \oddsidemargin
\begin{document}
\topmargin -.6in
\def\br{\begin{eqnarray}}
\def\er{\end{eqnarray}}
\def\be{\begin{equation}}
\def\ee{\end{equation}}
\def\nn{\nonumber}
\def\({\left(}
\def\){\right)}
\def\a{\alpha}
\def\b{\beta}
\def\d{\delta}
\def\D{\Delta}
\def\eps{\epsilon}
\def\g{\gamma}
\def\G{\Gamma}
\def\h{ {1\over 2}  }
\def\hp{ {+{1\over 2}}  }
\def\hm{ {-{1\over 2}}  }
\def\k{\kappa}
\def\l{\lambda}
\def\L{\Lambda}
\def\m{\mu}
\def\n{\nu}
\def\o{\over}
\def\O{\Omega}
\def\p{\phi}
\def\rh{\rho}
\def\s{\sigma}
\def\t{\tau}
\def\th{\theta}
\def\ii {\'\i  }
\def\pa{\partial}
\begin{titlepage}
\begin{center}
{\large {\bf Ladder operators for subtle hidden shape invariant potentials}} 
\footnotemark 
\footnotetext{PACS No. 03.65.Fd, 11.30.Pb, 31.15.Pf - Key words: Supersymmetric quantum mechanics, shape invariance, ladder operators} 
\end{center} 
\normalsize 
\vskip 1cm 
\begin{center} {\it  Elso Drigo Filho} 
\footnotemark \footnotetext{elso@df.ibilce.unesp.br}  \footnotemark \footnotetext{Work
partially supported by CNPq} \\ 
Instituto de Bioci\^encias, Letras e  Ci\^encias Exatas,
IBILCE-UNESP\\ 
Rua Cristov\~ao Colombo, 2265 -  15054-000 S\~ao Jos\'e do Rio Preto - SP\\ 
\vskip 1cm 
{\it Regina Maria  Ricotta } \footnotemark \footnotetext{regina@fatecsp.br}\\ Faculdade de
Tecnologia de S\~ao Paulo, FATEC/SP-CEETEPS-UNESP \\ 
Pra\c ca  Fernando Prestes, 30 -  01124-060 S\~ao Paulo-SP\\ Brazil\\ 
\vskip 1cm 
 
{\bf  Abstract}\\  
\end{center}
\par \vskip .3in \noindent

Ladder operators can be constructed for all potentials that present the integrability  condition
known as shape invariance, satisfied by most of the exactly solvable potentials. Using the
superalgebra of supersymmetric quantum mechanics  we construct the ladder operators for two
exactly solvable potentials that present a subtle hidden shape invariance.\\

\end{titlepage}

\noindent {\bf 1. Introduction} \\
Two decades ago it was shown that a subset of the exactly solvable  potentials share an 
integrability condition characterized by a reparametrization invariance known as shape
invariance, \cite {Gedenshtein}. In other words, not all the exactly solvable potentials seem to
be shape invariant, a property introduced within the concept of supersymmetric
quantum mechanics, SQM, \cite{Cooper}.  Ten  years ago  it was shown that this shape invariance
condition has an underlying algebraic structure associated to  Lie algebras, \cite{Aizawa}-\cite{Balantekin2}. For the potentials that share this property it
is possible to define coherent states and  ladder operators  defined in terms of
the bosonic operators of the superalgebra, similar to the harmonic oscillator ladder operators.

Here our interest is in two particular exactly solvable potentials that, although known not to
share this property,  appear to have a subtle shape invariance, hidden by a special choice of the
parameters of the transformation. In what follows we introduce the general formulation to
construct the ladder operators and then apply the methodology to two different exactly
solvable potentials that do not present shape invariance at first,  the case of the free
particle confined in a box and the case of the Hulth\'en potential. 

Consider a system described by a given potential $V$. The associated Hamiltonian $H$ can be
factorized in terms of  bosonic operators and its lowest energy state, in $\hbar = c = 1$ units,
\cite{Cooper}, 
\be H_+ = H - E_0 =  -{d^2 \o d x^2} + V_+(x) =  A^+A^-   \ee 
where $ E_0$ is the lowest eigenvalue.  The bosonic operators are  defined in terms of the 
superpotential $W(x, a)$, which is a function of the position variable and a set of parameters,
$a$, that represent space-independent properties of the original potential $V(r)$
\be 
\label{A_pm}
A^{\pm} =  \left(\mp {d \o dx} + W(x, a) \right) .
\ee 
\be H_+ = -{d^2 \o d x^2} + W^2(x, a) - W'(x, a)     \ee 
The partner Hamiltonian of $H_-$ is given by
\be H_- = A^-A^+  = -{d^2 \o d x^2} + V_-(x)  \ee 
\be \label{H_-} 
H_- = -{d^2 \o d x^2} + W^2(x, a) + W'(x, a)     \ee 
The Hamiltonians $H_+$ and $H_-$ have the same spectra except for the ground state of $H_+$, for
which there is no corresponding state in the spectra of $H_-$.  As a consequence of the
factorization of the Hamiltonian $H$, the Riccati equation must be satisfied, 
\be
\label{Riccati} W^2(x, a) - W'(x, a)=  V(x) - E_0 = V_+. 
\ee 
and the corresponding potential $V_-(x)$ satisfies
\be  W^2(x, a) + W'(x, a) = V_-(x).
\ee
The shape invariant condition states that 
\be
\label{SI}
V_-(x,a_0) - V_+(x,a_1) = R(a_1)
\ee
where $R(a_1)$ is independent of any dynamical variable and $a_1 = f(a_0)$. In terms of the
bosonic operators the above condition is given by
\be
A^-(a_0) A^+(a_0) - A^+(a_1) A^-(a_1) = R(a_1).
\ee
Potentials that
satisfy this condition are exactly solvable. The contrary is not true: an exactly solvable
potential may not be shape invariant.  In this work, we consider the shape invariance involving
translations of the parameters a:
\be
a_1 = a_0 + \eta
\ee
where $\eta$ is the translation step. Thus it is possible to define operators $T(a_0)$ as
\be
T(a_0) = exp(\eta  {\pa \o \pa a_0})
\ee
and 
\be
T^{-1}(a_0) = T^{\dagger}(a_0) =exp(-\eta  {\pa \o \pa a_0}).
\ee
These operators only act on objects defined in the parameters space.

Now we introduce the ladder operators, such as the creation and annihilation operators, by composing
the translation operators $T$ and the bosonic operators $A^{\pm}$,
\br
\label{ladder}
B_+(a_0) = A^+(a_0) T(a_0) \nonumber \\
B_-(a_0) = T^{\dagger}(a_0) A^-(a_0).
\er
The operators $B_{\pm}$ present the necessary algebraic structure \cite{Balantekin} to identify them
as ladder operators.  As such they are analogous to the harmonic oscillator ladder operators
\be
H_+ = A^+A^-   = B_+B_-   .
\ee
Thus, the ground state  must obey
\be
B_-(a_0) \Psi_0(x, a_0) = A^-(a_0) \Psi_0(x, a_0) = 0
\ee
or equivalently
\be
\Psi_0(x, a_0)  = N exp( -\int_0^x W(\bar x) d\bar x).
\ee
The excited states are obtained by the repeated action of the creation operator on the ground
state
\be
\Psi_n(x, a_0)  = (B_+)^n (a_0)\Psi_0(x, a_0) 
\ee

At this point we emphasize that this algebraic approach is self-consistent and it allows us to
determine the energy eigenvalues and eigenfunctions of a bound-state Schr\"odinger equation from
supersymmetric and shape invariance properties of the system. The energy is given by
\be
\label{energy}
E_n = E_0 + \sum_{k=1}^n R(a_k).
\ee
\\
\noindent {\bf 2. The free particle in a box}\\ 

Consider the case of a free particle confined in a box of infinite walls. The potential
is written as 
\br V(x) &=& 0 \;,\;\;\;\; 0 < x < \pi \nonumber \\
&= & \infty\;,\;\; -\infty < x < 0 \;\;\;;\;\; x > \pi
\er
and the factorised Hamiltonian in this case is \cite{Cooper}
\br
\label{H_+}
H_+ &=& H - E_0 = - {d^2\o dx^2}  + V_+(x)  \nonumber \\
&=& - {d^2\o dx^2}  - 1
\er
where $H$ is the original Hamiltonian with ground state energy eigenvalue $E_0 = -1$ so
that the ground state of $H_+$ is zero. The superpotential that factorises $H_+$
is 
\be
W(x) = - cot(x)
\ee
and its supersymmetric partner is 
\br
\label{H_-}
H_- &=& - {d^2\o dx^2} + V_-(x)\nonumber \\
&=& - {d^2\o dx^2} + {2 \o sin^2(x)} - 1.
\er
At this point we recollect the general form for the superpotential of the hierarchy
\cite{Sukumar}
\be
\label{Wn}        
W_n(x) = - n\; cot(x) 
\ee
where $n$ is a natural number different from  zero, ($ n=1,2,3...$). The
hierarchy is such that $E_n^{(1)}=n^2$ and the $n$-th member  of the  
super-family potential is
\be                 
V_n(x) - E_0^{(n)} =  {n(n-1) \o sin^2(x)} - n^2.
\ee   
Thus, it is not shape invariant since $V_+ = -1$ and
$V_- = {2 \o sin^2(x)}-1$. 
However,  inspired by the superpotential of the hierarchy, equation(\ref{Wn}), we rewrite the
superpotential  of $H_+$ in terms of a parameter $a_0$,  
\be
\label{W1a_0}
W(x, a_0) = - a_0 cot (x). 
\ee
The superpotential (\ref{W1a_0}) is a special case of the more general Infeld-Hull type E  potential \cite{Infeld}, whose shape invariance has been discussed in \cite{Aizawa}. The related Hamiltonian is given by 
\br
\label{H_+1 SI}
H_+ &=& - {d^2 \o dx^2} + V_+ \nonumber \\
&=& - {d^2 \o dx^2}
 + {a_0(a_0 - 1) \o sin^2(x)} - a_0^2.
\er
Its supersymmetric partner is given by
\br
\label{H_-1 SI}
H_-  &=& - {d^2 \o dx^2} + V_- \nonumber \\
&=& - {d^2 \o dx^2} + {a_0(a_0 + 1) \o sin^2(x) }- a_0^2.
\er
Thus, setting $a_0 = 1$ we recover $H_{\pm}$ of the free particle given by equations (\ref{H_+})
and (\ref{H_-}). Now we can test the shape invariance. Substituting the potentials of equations
(\ref{H_+1 SI}) and (\ref{H_-1 SI}) into (\ref{SI}) we obtain the following expression:
\be
R(a_1) = ({a_0(a_0 + 1) \o sin^2(x) }- a_0^2) - ({a_1(a_1 - 1) \o sin^2(x) }- a_1^2)
\ee
which is an x-independent for $a_1 = a_0 + 1$. The step is then $\eta = 1$ and thus
\be R(a_1) = a_1^2 -
a_0^2 = 2a_0 + 1.
\ee
The other steps shall be given by $a_k = a_0 + k$ and 
\br
R(a_k) &=& a_k^2 - a_{k-1}^2  \nonumber \\
&=& (a_0 + k \eta)^2 - (a_0 + (k-1) \eta)^2 \nonumber \\
&=& 2k +1
\er
where we have set $a_0 = 1$. The energy levels, evaluated from equation (\ref{energy}) will be
given by
\br
E_n &=& E_0 + \sum_{k=1}^n R(a_k) \nonumber \\ 
&=& 1 + \sum_{k=1}^n (2k +1) \nonumber \\ 
&=& (n +1)^2 
\er
as expected.
The ladder operators, evaluated from equations (\ref{A_pm}) and (\ref{ladder}) and the
superpotential (\ref{W1a_0}) are then given by
\be
B_+(a_0) = \left ( -{ d \o dx} - a_0 cot(x) \right )exp(- {\pa \o \pa a_0})
\ee
and 
\be
B_-(a_0) = exp(- {\pa \o \pa a_0}) \left ( { d \o dx} - a_0 cot(x) \right )
\ee
and from the fact that 
\be
B_-(a_0) \Psi_0 (x, a_0) = 0   
\ee
we arrive at
\be
\Psi_0 (x, a_O) \propto (sin x)^{a_0} = sin x\,,\,\,\,\, a_0 = 1
\ee
the ground state of the starting   Hamiltonian. The excited states are constructed through the action of $B_+(a_0)$ in the ground state. For  the
first excited state we have
\be
\Psi_1(x,a_0) \propto B_+(a_0)\Psi(x,a_0) = -(2a_0 +1) cos x (sin x)^{a_0} 
\ee
and this gives 
\be
\Psi_1(x,a_0) \propto -{3\over 2} sin 2x \,,\,\,\,\,a_0 = 1 
\ee
which is correct apart from a normalisation factor.\\

\noindent {\bf 3. The Hulth\'en Potential}\\ 

The Hulth\'en Potential, in atomic units, is given by:
\be
\label{Potential}
V_H(x) = - {2 \d e^{-\d x}\o 1-e^{-\d x}} 
\ee
where $\d$ is the screening parameter. From early results, \cite{Drigo}, the partner
Hamiltonians are given by
\br
\label{H_+ Hulthen}
H_+ &=& H - E_0 =  - {d^2 \o dx^2}  + V_+ \nonumber \\
&=&  - {d^2 \o dx^2} - {2 \d e^{-\d x}\o 1-e^{-\d x}} + (1 - {\d \o 2})^2
\er
and
\br
\label{H_- Hulthen}
H_- &=& - {d^2 \o dx^2}  + V_- \nonumber \\
&=& - {d^2 \o dx^2} -  {2 \d  e^{-\d x}\o 1-e^{-\d x}} +  {2{\d}^2 e^{- \d x}\o 
(1-e^{-\d x})^2}  + (1 - {\d \o 2})^2
\er
where $H$ is the Hamiltonian of the original problem and the superpotential is given by
\be
W(x) = - { \d e^{-\d x}\o 1-e^{-\d x}} + 1 - {\d \o 2}.
\ee
From equations (\ref{H_+ Hulthen}) and (\ref{H_- Hulthen}), we see that  
$V_+$ and $V_-$ surely are not shape invariant since we cannot satisfy equation ({\ref{SI}) by a
change of parameters between them. At this point, however, we address ourselves to the results
concerning the $n$th member of the Hulth\'en hierarchy \cite{Drigo}.  The superpotential is
given by 
\be
W_n(x) = - {n\d e^{-\d x}\o 1-e^{-\d x}} + {1\o n} - {n\o 2}\d
\ee
which corresponds to the $n$th member of the Hamiltonian hierarchy with the potentials
\br
\label{potential}
V_n(x)- E_0^{(n)} &=& W_n^2(x) -{d\o dr}W_n(x)  \nonumber \\
&=& {n(n-1){\d}^2 e^{-\d x}\o (1-e^{-\d x})^2} - { 2\d e^{-\d x}\o
1-e^{-\d x}} + ({1\over n} - {n\over 2} \d )^2
\er
Notice that $V_+$ corresponds to $n=1$. 
Thus, again inspired by the hierarchy superpotential we suggest writing the
superpotential in terms of a quantity $a_0$ such that
\be
\label{W2a_0}
W(x, a_0) = - {a_0 \d e^{-\d r}\o 1-e^{-\d r}} + {1\o a_0} - {a_0\o 2}\d 
\ee
and evaluate the Hamiltonian 
\br
\label{H_+ SI}
H_+ &=& (-{ d \o dx} + W(x, a_0)) ({ d \o dx} + W(x, a_0)) = - {d^2 \o dx^2}  + V_+(x,
a_0)\nonumber \\
 &=& - {d^2 \o dx^2} - {2 \d  e^{-\d x}\o 1-e^{-\d x}} + {a_0(a_0 -1) {\d}^2 e^{- \d x}\o
(1-e^{-\d x})^2} + ({1\o a_0}- {a_0 \d \o 2})^2 .
\er
Its supersymmetric partner is given by
\br
\label{H_- SI}
H_- &=& ({ d \o dx} + W(x, a_0)) (-{ d \o dx} + W(x, a_0)) = - {d^2 \o dx^2}  +
V_-(x, a_0)
\nonumber \\
&=& - {d^2 \o dx^2} - {2 \d  e^{-\d x}\o 1-e^{-\d x}} + {a_0(a_0 + 1) {\d}^2 e^{- \d x}\o
(1-e^{-\d x})^2} + ({1\o a_0}- {a_0 \d \o 2})^2. 
\er
Thus, setting $a_0 = 1$ we recover $H_+$ and  $H_-$ of the original Hulth\'en problem given by
equations (\ref{H_+ Hulthen}) and (\ref{H_- Hulthen}). 

Now we can test the shape invariance.  Substituting the potentials of equations (\ref{H_+ SI})
and (\ref{H_- SI}) into (\ref{SI}) we obtain an x-independent expression,
\be
R(a_1) = ({1\o a_0} - {a_0 \d \o 2})^2 - ({1\o a_1} - {a_1 \d \o 2})^2
\ee
for $a_1 = a_0 + 1$. The step $\eta = 1$.  The other steps
shall be given by $a_k = a_0 + k$. This gives
\be
R(a_k) = ({1\o a_{k-1}} - {a_{k-1} \d \o 2})^2 - ({1\o a_k} - {a_k \d \o 2})^2 
\ee
which results in 
\be
R(a_k) = {1 + 2k \o k^2 (k^2+1)} - (2k + 1) {{\d}^2 \o 4}.
\ee
when we set $a_0 = 1$. The energy levels are then  given by
\br
E_n &=& E_0 + \sum_{k=1}^n R(a_k) \nonumber \\ 
&=& - \left({1\o n+1} - (n+1) {\d \o 2}\right)^2 \;\;,\;\;\; n=0,1,2,...
\er
as expected.\\
The ladder operators, evaluated from equations (\ref{A_pm}) and (\ref{ladder}) and the
superpotential (\ref{W2a_0}), are then given by
\be
B_+(a_0) = \left ( -{ d \o dx} - {a_0 \d e^{-\d x}\o 1-e^{-\d x}} + {1\o a_0} - {a_0\o 2}\d 
\right )exp( {\pa \o \pa a_0})
\ee
and 
\be
B_-(a_0) = exp(- {\pa \o \pa a_0}) \left ( { d \o dx} - {a_0 \d e^{-\d x}\o 1-e^{-\d x}} + {1\o
a_0} - {a_0\o 2}\d  \right )
\ee
and from the fact that 
\be
B_-(a_0) \Psi_0 (x, a_0) = 0   
\ee
we arrive at
\be
\Psi_0 (x, a_0) = exp\left(-({1\o a_0} - {a_0 \d \o 2})x\right) \left( 1-e^{-\d x}
\right)^{a_0} 
\ee
the ground state of the starting   Hamiltonian. The excited states are constructed through the
action of $B_+(a_0)$ in the ground state. The first excited state is given by
\be
\Psi_1(x,a_0) \propto B_+(a_0)\Psi(x,a_0)  = A_+(a_0)\Psi(x,a_0 + 1)
\ee
and this gives, for $a_0=1$,
\be
\Psi_1(x) \propto {3 \o 2} exp\left( -({1 \o 2} - \d) x \right) ( 1-e^{-\d x}) \left( (1-
\d) -  e^{-\d x}( 1+ \d) \right)
\ee
as expected.\\

\noindent {\bf 4. Conclusions} \\
We have shown that two different exactly solvable potentials present a hidden shape invariance,
which is seen after implementing a parameter which will develop the required transformation. 
The  implementation of this parameter, which is in fact fixed and equal to the unity in both
cases, enables the construction of the ladder operators analogous to the creation and
annihilation operator of the harmonic oscillator case.

We notice that Hulth\'en potential, written in terms of hyperbolic functions, is a particular
case of the  exactly solvable and shape invariant Eckart potential for fixed values of the
parameters. However this is not obvious at first, starting from the factorisation of the
Hulth\'en potential and testing the shape invariance condition.  The same argument is valid to
the case of the free particle confined in a box and its relation to the Rosen-Morse I
potential.\\

\noindent {\bf 5. Acknowledgements} \\
The authors would like to thank Professors S. Salam\'o,  S. Codriansky, A. Khare and U. P.
Sukhatme for useful comments and discussions.

\end{document}